\documentclass[manuscript]{aastex631}

\usepackage{comment}
\usepackage{amsmath}
\usepackage{amssymb}
\usepackage{hyperref}
\usepackage[caption=false]{subfig}
\newcommand{\pkdgrav}{\texttt{pkdgrav}}

\begin{document}

\title{An efficient numerical approach to modeling the effects of particle shape on rubble-pile dynamics}

\correspondingauthor{Julian C. Marohnic}
\email{jmarohni@umd.edu}

\author[0000-0002-6810-7491]{Julian C. Marohnic}
\affiliation{Department of Astronomy \\
University of Maryland \\
College Park, MD 20742, United States}

\author[0000-0003-4396-1728]{Joseph V. DeMartini}
\affiliation{Department of Astronomy \\
University of Maryland \\
College Park, MD 20742, United States}

\author[0000-0002-0054-6850]{Derek C. Richardson}
\affiliation{Department of Astronomy \\
University of Maryland \\
College Park, MD 20742, United States}

\author[0000-0003-4045-9046]{Yun Zhang}
\affiliation{Department of Aerospace Engineering \\
University of Maryland \\
College Park, MD 20742, United States}

\author[0000-0002-0906-1761]{Kevin J. Walsh}
\affiliation{Southwest Research Institute \\
Boulder, CO 80302, United States}

\begin{abstract}
We present an approach for the inclusion of non-spherical constituents in high-resolution \textit{N}-body discrete element method (DEM) simulations. We use aggregates composed of bonded spheres to model non-spherical components. Though the method may be applied more generally, we detail our implementation in the existing \textit{N}-body code \pkdgrav{}. It has long been acknowledged that non-spherical grains confer additional shear strength and resistance to flow when compared with spheres. As a result, we expect that rubble-pile asteroids will also exhibit these properties and may behave differently than comparable rubble piles composed of idealized spheres. Since spherical particles avoid some significant technical challenges, most DEM gravity codes have used only spherical particles, or have been confined to relatively low resolutions. We also discuss the work that has gone into improving performance with non-spherical grains, building on \pkdgrav{}'s existing leading-edge computational efficiency among DEM gravity codes. This allows for the addition of non-spherical shapes while maintaining the efficiencies afforded by \pkdgrav{}'s tree implementation and parallelization. As a test, we simulated the gravitational collapse of 25,000 non-spherical bodies in parallel. In this case, the efficiency improvements allowed for an increase in speed by nearly a factor of three when compared with the naive implementation. Without these enhancements, large runs with non-spherical components would remain prohibitively expensive. Finally, we present the results of several small-scale tests: spinup due to the YORP effect, tidal encounters, and the Brazil-nut Effect. In all cases, we find that the inclusion of non-spherical constituents has a measurable impact on simulation outcomes.
\end{abstract}

\section{Introduction} \label{sec:intro}
Most small solar system objects are believed to be loose, unconsolidated masses of material, rather than monolithic bodies \citep{walsh2018rubble}. These ``rubble piles" are largely held together by gravity and may be of various composition depending on their location in the solar system and specific history. Substantial work has been devoted to studying rubble-pile bodies, including via numerical techniques. Notable among these numerical methods are smoothed-particle hydrodynamics (SPH) \citep{benz1995simulations,jutzi2015sph} and discrete-element method (DEM) codes \citep{richardson1998tidal,sanchez2012simulation,schwartz2012implementation,sanchez2014strength,zhang2020tidal}. In this study, we  focus exclusively on DEM, which treats rubble-pile constituents as separate elements explicitly rather than as a continuum via approximate constitutive relations. In particular, we will consider the effects of the \textit{shapes} of the granular elements and the effects of these shapes on the behavior of the body as a whole. 

\cite{solem1994density} and \cite{asphaug1996size} used a frictionless, hard-sphere particle model to estimate the size and density of comet Shoemaker-Levy 9 after its tidally induced disaggregation. \cite{movshovitz2012numerical} approached the same problem using polyhedral particles, determining that disruption is more difficult than in the case of a similar sphere-based model. However, the study was restricted to relatively low-resolution simulations (about 4000 grains) due to computational limitations. \cite{walsh2006binary,walsh2008steady} used a hard-sphere numerical model to study the formation of binary asteroid systems via tidal disruption of rubble-pile asteroids and found that this mechanism alone was not sufficient to explain the observed binary fraction in the near-Earth asteroid population. The authors later used the same model to show that YORP spin-up of rubble piles can create fast-rotating asteroids with close secondaries \citep{walsh2008rotational}. More recent work has applied the soft-sphere discrete element model \citep[SSDEM;][]{cundall1979discrete}---which allows for persistent particle contacts and friction---to the problem of disruption and shape change of rubble piles via spin-up and tidal forces, though these studies are still largely confined to spherical particles \citep{sanchez2012simulation,walsh2012spin,yu2014numerical,zhang2018rotational}. Nevertheless, constituent shape is believed to be an important factor in the behavior of granular media generally and should be considered in the context of small rubble-pile bodies as well.

\subsection{Importance of particle shape in granular media}

Granular media made up of non-spherical components exhibit higher shear strength than those composed of spheres (see \cite{wegner2014effects} and references therein). In a compacted state, non-spherical particles in a granular medium will interlock with each other. When subjected to shear, these interlocked constituents cannot easily slide past one another without the entire medium first expanding, or ``dilating." The overall tendency of a granular body to dilate under shear force is likewise known as ``dilatancy." Though they are not monolithic objects, rubble-pile asteroids are subject to confining pressure due to their own self-gravity. As a result, when they experience shearing from a tidal encounter or spin-up that exceeds their shear strength, the particles that make up the body will tend to move relative to each other as in any other granular medium under compression.

While the spherical particles in traditional DEM models can easily slide or roll past one another, this is more difficult for particles with other shapes due to the increased dilatancy of the medium. Thus, we expect that the physical rubble piles in the solar system that are made up of irregular constituents will have a higher effective shear strength and will be more difficult to disrupt than in idealized DEM simulations with spherical particles. A related effect has been documented in the case of terrestrial granular flows in both experiments \citep{sarkar2019interpretation} and DEM simulations \citep{cleary2008effect,mead2015validation}. When a granular solar system body is disrupted, the resulting fragments and reaccumulated successor bodies may have different properties from those in the spherical-grain case. They are also likely to be able to maintain shapes further from their fluid-equilibrium shapes. These differences have implications for the understanding of small-body internal structure. In reality, small solar system bodies are not made up of spheres, whether on the scale of small grains or large boulders, as imagery from the recent \textit{Hayabusa2} and \textit{OSIRIS-REx} missions, for example, have made clear \citep{michikami2019boulder,walsh2022assessing}. Since many of the small bodies in our solar system are subjected to stresses from spin-up and tidal encounters at some point during their evolution, it is important to understand and quantify the effects of irregular particle shapes. For example, what role does particle shape play in setting the critical spin limit for rubble piles, and to what extent? How does particle shape affect binary formation under tidal and rotational stresses? Given the resolution and efficiency we can now achieve with non-spherical particles in the code we describe here, these are questions we can begin to answer.

Most DEM codes used in the context of small solar system bodies have relied on spherical particles for the simplicity and computational efficiency they afford. That said, some N-body DEM codes have included non-spherical particles of various construction, with most examples either using poly-ellipsoids \citep{roig2003interacting}, or polyhedra \citep{ferrari2020role,sanchez2021contact}. Due to the complexity of applying contact physics and interparticle gravity to irregular shapes, these efforts have historically had to compromise on the fidelity of the physics model, the number of particles included in simulations, or both. \cite{ferrari2020role} include polyhedral particles with soft-sphere contacts in their model, but don't allow for gravitational torques as they treat these polyhedra as point particles when calculating gravitational interactions. And while this method uses a tree to reduce the cost of gravity calculations, it is confined to a single GPU, ultimately leading to memory limitations on overall resolution. \cite{sanchez2021contact} also allow for non-spherical, self-gravitating particles, but use a non-smooth contact dynamics method rather than an SSDEM approach and likewise omit gravitational torques. Alternately, there are a number of existing DEM codes that are capable of modeling the interaction of hundreds of thousands or even millions of non-spherical particles with full soft-sphere DEM contacts \citep{zhao2006three,knuth2012discrete,longmore2013towards,nguyen2019aspherical}. However, none of these include the expensive calculations of interparticle gravitational forces that are critical for studying the dynamics of rubble-pile bodies. This work presents a scheme for assembling spheres into rigid, non-spherical grains. Although this implementation leverages our code’s existing ability to quickly compute gravity and contact interactions between large numbers of particles, the techniques described here are broadly useful to the study of small, rubble-pile bodies and could be applied in other N-body DEM codes. We use the ``glued-sphere" method popular in granular dynamics \citep{nolan1995random,song2006contact}, attaching the existing spherical \pkdgrav{} particles together in arbitrary shapes. This approach seamlessly joins the existing soft-sphere physics model and the highly efficient gravity tree from \pkdgrav{}, allowing for N-body simulations with non-spherical particles, gravitational and collisional torques between grains, and resolutions up to hundreds of thousands or more particles. To the best of our knowledge, this represents the fastest implementation that combines all of these features. Section \ref{sec:methods} describes the basic computational methods while Section \ref{sec:efficiency} details how we have improved the efficiency of our particular implementation. We present some applications of the new code in Section \ref{sec:applications} as proofs of concept and to lay the groundwork for future projects. Section \ref{sec:conclusion} provides a brief summary of this work.

\section{Computational Methods} \label{sec:methods}

We begin with a brief review of our numerical approach to modeling particle contacts with SSDEM in our code \pkdgrav{} prior to implementing non-spherical shapes. This is followed by details on the modifications needed for modeling non-spherical constituents.

A note on our terminology: instead of constructing polyhedral particles with flat faces and edges, the non-spherical particles or grains that we wish to model are made up of spheres locked together. The spheres themselves are the object traditionally considered ``particles" in \pkdgrav{} and are analogous to the particles used in most other DEM codes. In describing our implementation, we must frequently refer to the individual spherical particles that already exist in \pkdgrav{}, as well as the rigid collections of these spheres that form the non-spherical grains. To avoid confusion, we will refer to the non-spherical assemblies as ``aggregates" or ``bonded aggregates." The word ``particle" left unmodified should be assumed hereafter to refer to the spherical elements that make up the aggregates. In addition, ``constituents," ``components," or ``grains" will refer to either spherical particles or bonded aggregates in the context of the pieces that compose a rubble-pile body, regardless of their shape or size. Since most force calculations and position and velocity evolutions in our implementation are carried out at the particle level, this distinction is important.

\subsection{Existing soft-sphere implementation for spherical particles
}
\label{subsec:soft-sphere}

Our approach builds on the existing numerical gravity code \pkdgrav{} \citep{richardson2000direct,stadel2001cosmological}. \pkdgrav{} uses a hierarchical tree algorithm that reduces the cost of locating neighboring particles to an $\mathcal{O}(N\log{}N)$ operation, where $N$ is the number of particles. Interparticle gravity is also computed using a tree to speed up the calculations, by replacing $\mathcal{O}(N^2)$ sums over individual particles with multipole expansions of the gravitational potential contributed by small or distant cells. The multipole expansions are taken to hexadecapole order as a middle ground between speed and accuracy. Both neighbor searching and gravity calculation are also parallelized, allowing \pkdgrav{} to distribute the work across an arbitrary set of processors for further speed optimization.

\pkdgrav{} also includes a soft-sphere discrete-element-method (SSDEM) scheme for treating particle interactions, which is described in much greater detail in previous works \citep{schwartz2012implementation,zhang2018rotational}. Unlike hard-sphere methods, SSDEM resolves collisions temporally, allowing particles to interpenetrate slightly as a proxy for surface deformation. \texttt{pkdgrav}'s SSDEM implementation uses a spring-dashpot model, in which overlaps between neighboring particles are detected and normal and tangential restoring forces are then modeled as damped springs following Hooke's law. Spring and damping constants are user-adjustable, but in practice must be set carefully to capture the properties of the particular material being modeled and to maintain physically realistic behavior. To ensure that particle overlaps remain small, spring constant settings must account the masses, sizes, and speeds of all constituents. The spring forces capture the effects of deformation at particle contacts, while the damping forces capture the effects of kinetic friction. \pkdgrav{} uses this approach to track forces and torques from twisting, rolling, and sliding friction. Particle overlaps are tracked for as long as particles remain in contact, and reaction forces depend not only on the degree of overlap at the current time step but also on the contact history. SSDEM is a substantially more realistic model of the granular physics we are interested in than hard-sphere DEM, able to capture multiple simultaneous and persistent contacts self-consistently. These are exactly the sort of interactions at work in small rubble-pile bodies; using an SSDEM model thus allows us to more accurately simulate the interaction of systems made up of hundreds of thousands of particles in contact.

\pkdgrav{} uses a fixed-step, second-order leapfrog method to integrate gravity and contact interactions. The leapfrog integrator updates center of mass positions and velocities for both individual particles and aggregates, while the orientations and spins of bonded aggregates (described in greater detail below) are instead evolved with an adaptive Runge-Kutta method. While leapfrog integration is designed for equations of motion with no explicit velocity dependence, the friction model introduces velocity-dependent damping terms. To account for this, we also calculate ``predicted" velocities and spin vectors for each particle. Using predicted velocities technically reduces the scheme to first order, but because the velocity-dependent terms are all related to damping and because the time steps are very small in order to resolve the contacts, this is not an issue in practice \citep{schwartz2012implementation}.

\subsection{Bonded aggregates} \label{non-sph}

To address the question of grain shape in rubble-pile bodies, we need the ability to simulate non-spherical constituents. Our approach is to construct compound pseudo-particles, which we call ``bonded aggregates," that consist of multiple spherical particles ``glued" together with unbreakable bonds. This approach was used in the hard-sphere model of \pkdgrav{} \citep{richardson2009numerical} but requires updating for the soft-sphere model. We arrange arbitrary numbers of spherical particles in any desired shape and then fix their relative positions so that they behave as a unit, creating rigid, non-spherical aggregates. One of the primary advantages of constructing non-spherical constituents this way is that computing forces and torques is relatively straightforward. Since we already track the positions and velocities of the constituent spheres, we can easily calculate the total mass and COM positions and velocities of these bonded aggregates. They can then be treated as discrete objects that receive a unique identifying numerical tag for tracking. Much of the existing machinery for calculating contact and gravitational forces can also be reused in the context of bonded aggregates. In addition, we retain the static friction model used in the sphere-based version of \pkdgrav{}, which allows us to capture the effects of grain roughness in addition to grain shape. In this section, we describe in more detail how this method is implemented. To model a full granular assembly with many non-spherical constituents using \pkdgrav{} efficiently, further optimization is required to complement these modifications to our physics scheme (Section\ \ref{sec:efficiency}).

\begin{figure}
\centering\includegraphics[width=\linewidth]{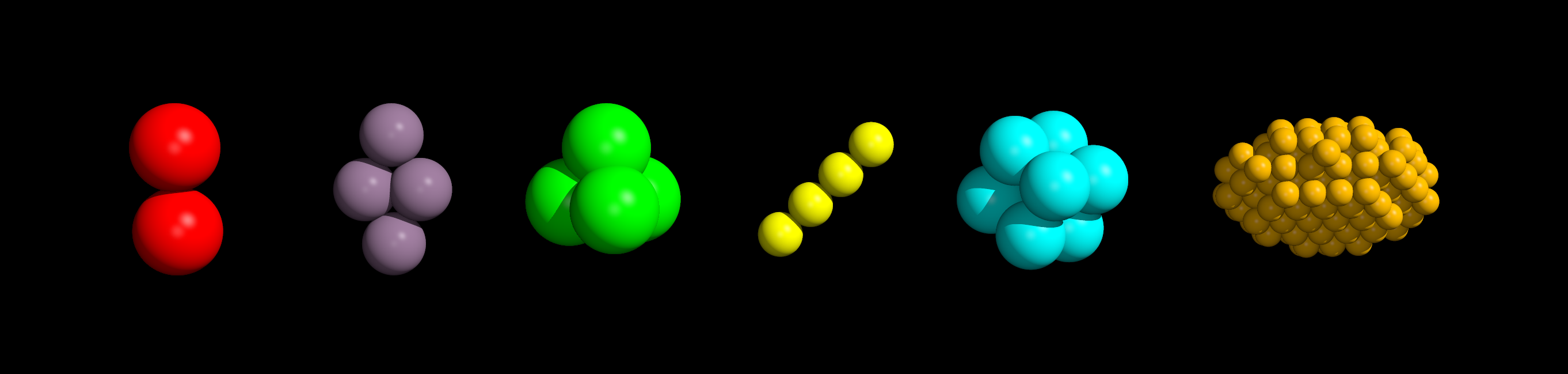}
\caption{Non-spherical constituents or ``bonded aggregates" as implemented in \pkdgrav{}. From left to right, a two-particle ``dumbbell" shape, a 4-particle planar diamond, a 4-particle tetrahedron, a 4-particle rod, an 8-particle cube, and a 155-particle irregular aggregate.}
\label{fig:aggs_demo}
\end{figure}

\subsubsection{Preliminaries}

We use the symbol $A$ as a subscript to refer to a bonded aggregate and the same symbol in the context of a summation represents the set of indices of all of the constituent particles of aggregate $A$. We define the total aggregate mass $M$, which is simply the sum of the masses of the constituent particles,
\begin{equation}
    \label{eq:agg_mass}
    M = \sum_{i\in A} m_i ,
\end{equation}
where $m_i$ is the mass of a particle in aggregate $A$. We also define the COM position $\textbf{r}_A$ of an aggregate, which can be considered the ``position" of the aggregate,
\begin{equation}
    \label{eq:agg_position}
    \textbf{r}_A = \frac{1}{M}\sum_{i\in A} m_i\textbf{r}_i,
\end{equation}
where $\textbf{r}_i$ is the COM position of a particle in the aggregate. Other aggregate-specific quantities will be introduced as needed.

While the COM positions and velocities of aggregates follow our previously established equations of motion, the introduction of bonded aggregates means we must account for rigid-body rotations as well. Aggregate rotations should obey the Euler rigid body equations, which we solve with a time-adaptive, fifth-order Runge-Kutta integrator. For a more detailed account of the integration scheme used for bonded aggregates, readers may consult \cite{richardson2009numerical}, which describes an earlier hard-sphere implementation of bonded aggregates in \pkdgrav{} using the same integration method.

\subsubsection{Predicted velocities}
\label{subsubsec:pred_vel}

We must calculate predicted velocities for bonded aggregates just as we did for spherical particles (see Sec.\ \ref{subsec:soft-sphere}). Consider an individual particle that is a member of a bonded aggregate. Since forces are tabulated on a particle-by-particle basis, the predicted velocity is still important in the context of bonded aggregates. In this case, to predict the movement of the particle we must take into account the rotation of the aggregate itself in addition to the current COM velocity of the particle. The predicted velocity estimate for a particle in an aggregate is now given by the following equation:
\begin{equation}
    \dot{\textbf{r}}_{i,n+1}^{\mathrm{pred}} =  \left[ \dot{\textbf{r}}_{A,n+\frac{1}{2}} + \frac{h}{2}\ddot{\textbf{r}}_{A,n} \right] + \left[ \boldsymbol\omega_{A,n+1} \times \left(\textbf{r}_{i,n+1} - \textbf{r}_{A,n+1} \right) \right]
\end{equation}
Calculating predicted velocities for particles belonging to aggregates necessarily works differently than in the case of stand-alone particles, since these particles are now part of rigid bodies. The first term in square brackets is the predicted COM velocity of the aggregate that particle $i$ belongs to, which is in turn imparted to the particle itself. The second term in square brackets represents the contribution of aggregate rotation to predicted particle velocity, determined by the spin rate of the aggregate and the distance of the particle from the center of the aggregate. We note that \textit{aggregate} spins need not be extrapolated to whole-number time steps in the same way that they are for individual particles, since aggregate spins and orientations are integrated separately from the primary leapfrog scheme with a Runge-Kutta method as described previously.

\subsubsection{Gravity}

Gravitational forces are calculated for each individual sphere in \pkdgrav{}. When the tree code is being used, which is typical, these force calculations are approximated. In the case of a self-gravitating assembly of independent spherical particles, we calculate the acceleration $\ddot{\textbf{r}}_i$ felt by each particle. In the case of an aggregate, we instead calculate the \emph{force} of gravity on each constituent particle in the aggregate $\textbf{F}_{i,g}$. We define this force as follows.
\begin{equation}
    \label{eq:particle_gravity}
    \textbf{F}_{i,g} = \sum_{j \neq i} 
    \frac{G m_i m_j \left(\textbf{r}_j - \textbf{r}_i \right)}
        {\left| \textbf{r}_j - \textbf{r}_i \right|^3}
\end{equation}
To find the net gravitational acceleration of a bonded aggregate $\ddot{\textbf{r}}_{A,g}$ we only need the vector sum of the gravitational forces on each particle in the aggregate, divided by the total aggregate mass.
\begin{equation}
    \label{eq:agg_gravity}
    \ddot{\textbf{r}}_{A,g} = \frac{1}{M}\sum_{i\in A}
    \textbf{F}_{i,g}.
\end{equation}
Individual spheres feel no net torque from gravity while a non-spherical aggregate in general will experience some torque $\textbf{N}_{A,g}$. The vector sum of the torque contributions from each constituent sphere in an aggregate is the net gravitational torque on the aggregate.
\begin{equation}
    \label{eq:gravity_torque}
    \textbf{N}_{A,g} = \sum_{i\in A} \left[ \left(\textbf{r}_i - \textbf{r}_A \right) \times      
                        \textbf{F}_{i,g} \right]
\end{equation}

\subsubsection{Contacts}

Particle contacts in general result in both a force and a torque on each particle. While gravitational forces act on particle centers, contact forces and torques act at the \emph{contact point}. In the case of a small aggregate composed of only a few particles, the difference in the effective lever arm can be quite significant. In the event of a contact, each particle feels a normal restoring force that depends on the extent of the overlap. If the particles have non-zero friction, they may also experience a tangential surface force. Contact forces between particles are then applied to the aggregates that they belong to, and the net center-of-mass acceleration on the aggregate due to contact forces $\ddot{\textbf{r}}_{A,c}$ is obtained according to
\begin{equation}
    \label{eq:agg_contact_accel}
    \ddot{\textbf{r}}_{A,c} = \frac{1}{M} \sum_{i\in A} m_i \ddot{\textbf{r}}_{i,c} ,
\end{equation}
where $\ddot{\textbf{r}}_{i,c}$ is the acceleration due to contact forces on constituent particle $i$. In other words, $m_i\ddot{\textbf{r}}_{i,c} = \textbf{F}_{i,N} + \textbf{F}_{i,T}$, where $\textbf{F}_{i,N}$ and $\textbf{F}_{i,T}$ are the normal and tangential contact forces (see \cite{schwartz2012implementation} for details).

The contact torque contribution to an aggregate from a constituent particle is calculated by considering the normal and tangential contact forces $\textbf{F}_{i,N}$ and $\textbf{F}_{i,T}$ felt by that particle. These forces act on a lever arm spanning from the aggregate center of mass to the contact point on the particle, rather than the particle center as in the case of gravity. To be precise, we say that the ``contact point" itself lies at the center of the circle formed by the intersection of the surface of the particle and its overlapping neighbor. The distance between the center of the particle and the contact point is equivalent to the lever arm $l_p$ for a spherical particle, which is given by
\begin{equation}
    \label{eq:lever_arm}
    l_p = \frac{s_p^2 - s_n^2 + \left|\textbf{d} \right|^2}{2\left|\textbf{d}\right|}.
\end{equation}
In Eq.\ \ref{eq:lever_arm}, $s_p$ and $s_n$ are the particle and neighbor radii, respectively, and $\textbf{d}$ is the distance between their centers. The vector that points from the center of the particle to the contact point is then given by $l_p \hat{\textbf{n}}$, where $\hat{\textbf{n}} = \textbf{d}/\left|\textbf{d}\right|$ is the unit normal vector pointing from the center of the particle toward the center of its neighbor. The vector sum of these torque contributions from each particle in the aggregate gives the net torque due to contact forces on a bonded aggregate in \pkdgrav{}:
\begin{equation}
    \textbf{N}_{A,c} = \sum_{i\in A} \left[ \left(l_p \hat{\textbf{n}} - \textbf{r}_A \right) \times m_i \ddot{\textbf{r}}_{i,c} \right].
\end{equation}
We note that the careful approach to applying gravity and contact forces described above is necessary for achieving proper physical behavior. Angular momentum conservation in particular degrades if torques are not applied at the contact points, especially in the case of large rubble piles.

\subsubsection{Rolling and twisting friction}
Rolling and twisting friction are currently included in \pkdgrav{} for spherical particles only, though a prescription for applying rolling and twisting friction to bonded aggregates is currently being developed and may be added in the future. There are certain circumstances in which this capability would enhance the realism of the code (e.g., a rod-shaped aggregate rolling on a flat surface). However, in the great majority of systems that are of interest to the field of small bodies, bonded aggregates would be used as constituents of a self-gravitating assembly rather than rolling or spinning freely on flat surfaces. Under these conditions, we expect that rolling and twisting friction will play a negligible role and, in fact, the non-spherical aggregate shapes should capture most of the physics that rolling and twisting friction schemes seek to parameterize in simulations with unbonded spheres.

\section{Computational Efficiency Improvements} \label{sec:efficiency}

In this section, we detail the methods of locating and operating on bonded aggregates in \pkdgrav{}, the improvements developed for these methods to complement the new physical model from Section \ref{non-sph}, and the profiling analysis that we employ to measure the efficiency of the new methods. References to processes occurring ``in serial" indicate that a single processor is handling the operations, while ``in parallel" means that the load is split across multiple processors, generally with each processor performing the same operation on a different subset of particles in the simulation. We emphasize that while the computational efficiency improvements detailed in this section are described in the context of \pkdgrav{}, these techniques are also applicable to any investigators intending to develop, modify, or improve other similar, parallel codes.

\subsection{Aggregate Constituent Searching} \label{search}
The original hard-sphere aggregate routines for \pkdgrav{} are optimized to handle a few large aggregates made up of many particles \citep{richardson2009numerical}. The routines keep track of aggregate properties (center-of-mass position/velocity, torques, etc.---see Table~\ref{tab:funcs}) in serial on a single leader processor while some or most of the constituent particle data may be distributed across multiple follower processors in groupings convenient for the interparticle gravity calculation. Load balancing performed each time step may cause these groupings to change. To carry out an operation on an aggregate, the leader processor requests data for each of the aggregate's constituent spheres, which may be on one or more follower processors. In the original routines, a ``brute-force" search finds particles bound in aggregates; it is a serial method that examines the particles on each processor one-by-one, eventually finding all particles contained within the aggregate and returning the information needed to calculate an aggregate property only after reaching the end of the full list of particles. If the number of aggregate particles is much larger than the number of aggregates and similar to the total number of particles in the simulation, each processor will have many particles belonging to a given aggregate, so the brute force search ``hits" frequently and works relatively well.

A noticeable inefficiency arises when the number of aggregates and the number of particles are both similar and large, as is the case when trying to simulate a granular system made up of many aggregates, each containing only a few constituent particles, like the aggregates in Fig.~\ref{fig:aggs_demo}. In this scenario, there are perhaps only two particles belonging to a given aggregate, but thousands of aggregates in the simulation. Every time the properties of a single aggregate require updates, the brute-force search wastes time scanning the whole particle list on each processor despite needing to find only two particles. Since the properties of each aggregate are updated every time step and each aggregate contains only a few particles, the originally tolerable brute-force search approaches an $O(N^2)$ process per step (where $N$ is the number of particles in the simulation); this is prohibitively expensive for large $N$ processes, like the simulations we perform in Section \ref{sec:applications}. The brute-force aggregate particle search is thus the main bottleneck for simulations involving large numbers of aggregate operations in \pkdgrav{}, and our aim in increasing the computational efficiency of these simulations is to improve the search method for identifying particles belonging to a given aggregate.

First, we solve the issue of efficiently locating aggregate particles by reordering all particles before the search begins. On each processor, particles are reordered by their aggregate index numbers---an identifying integer unique to each aggregate that is stored in the data structure of each of its particle members. All particles in a given simulation thus ``know" which aggregate they belong to, or whether they are simply unbound free particles. This reordering forces all particles already on a given processor into consecutive order by aggregate index, but allows for the possibility of aggregates that are split across processors. When data from aggregate member particles is required, we query each processor for its range of aggregate indices. If the index of the aggregate that we are operating on falls within the range bracketed by the end members on a processor, we search that processor for the relevant particles; if the desired aggregate index is outside a processor's range, it will not be queried. This prevents processors in parallel simulations from doing unnecessary work searching for particles that they will never find. However, this still leaves the issue of the $O(N^2)$ scaling of the brute-force search in serial simulations, as well as in parallel simulations when a processor does contain desired constituents in its range.

To further optimize the search method, we replace the original brute-force search with a binary-search algorithm \citep{bentley1975multidimensional} that exploits the new particle order and scales as $O(N\log N)$ to reliably and efficiently find the first member particle on a processor, and add a ``cache-line" method with best-case $O(N)$ scaling after the first member particle is found. The binary search is used whenever the particle information is not immediately known, i.e., not in the cache. The cache line stores identifying information of the previously found particle in order to easily step forward and find the next particle that needs to be acted on when the code returns to the function. This is a linear operation in $N$, so it scales well, and the cost of reordering by aggregate index at each time step is relatively small.

\subsection{Profiling Analysis} \label{prof_method}

In order to measure the increase in efficiency gained from the improvements described in Section \ref{search}, we used code profiling to determine the length of time \pkdgrav{} spends performing certain operations. In our analysis, we used the GPROF hybrid instrumentation and sampling profiler \citep{graham1982gprof}, one of the longest-standing profiling tools for compiled code. The output of our analysis contains a list of all the function calls in decreasing order of time spent in each function for a simulation that was run to completion. For each function, the number of calls to that function, the number of seconds spent in that function, and the corresponding percentage of the overall execution time devoted to that function are reported. The default resolution of the profiler is $0.01$~s and $0.01\%$, below which it will not report the actual time or percentage of time spent in a given function. The table in Appendix~\ref{app:AggFuncTable} lists the \pkdgrav{} functions modified to improve efficiency and gives brief descriptions of their purposes. Previously, these functions were the most time-consuming operations in simulations containing large numbers of aggregates relative to the total number of particles. In typical simulations of $N = 10,000$ equal-sized particles composing between 1,250 and 5,000 aggregates, these functions collectively account for $\geq$ 50\% of the total simulation time.

\subsection{Performance Scaling} \label{profiling}

\begin{figure}
\gridline{\fig{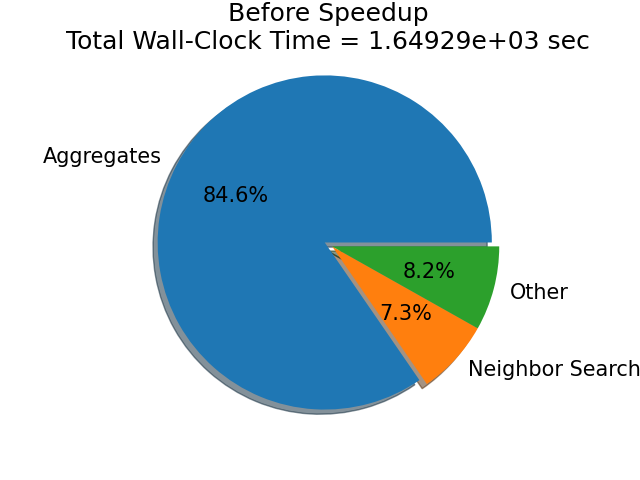}{0.45\textwidth}{(A)}
          \fig{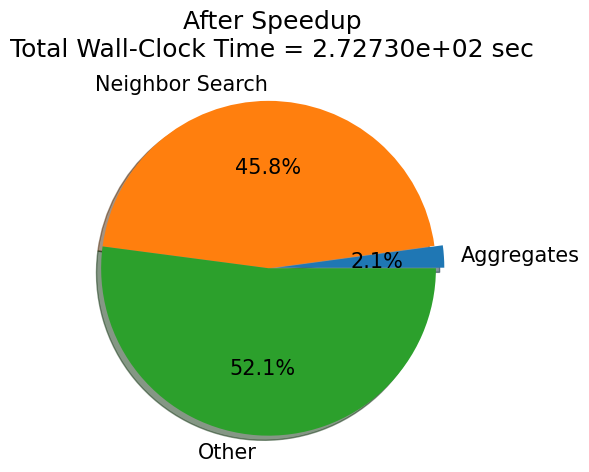}{0.42\textwidth}{(B)}}
\caption{A representative example of profiling results from the uniform gravity trials described in Section \ref{profiling}. The charts show the fractional simulation time spent on each of the three most computationally expensive sets of calculations both before (A) and after (B) the sort and search method enhancements. For uniform gravity tests like the one shown here, those categories are aggregate operations (blue), neighbor search operations (orange), and all other operations (green). Chart sectors are labeled accordingly. This example was taken from the set of box-filling simulations with $N =$ 6,400 member particles bound into a mix of different aggregate shapes and subjected to uniform lunar gravity. Notably, the percentage of simulation time spent in aggregates decreases from 84.4\% to 2.1\% with the new search and sort methods. The length of time spent on the neighbor search and ``Other" functions is constant between (A) and (B), but the total simulation time decreases in (B) (see Fig.~\ref{fig:profiling-tots}) due to the decrease in time spent on aggregate operations.}
\label{fig:profiling-unifgrav}
\end{figure}

\begin{figure}
\gridline{\fig{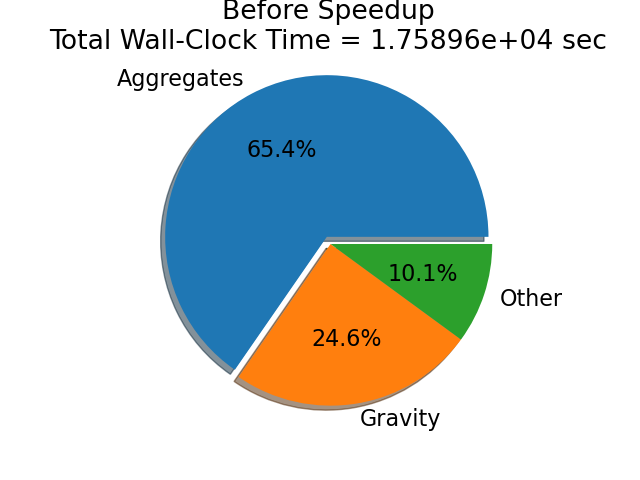}{0.45\textwidth}{(A)}
           \fig{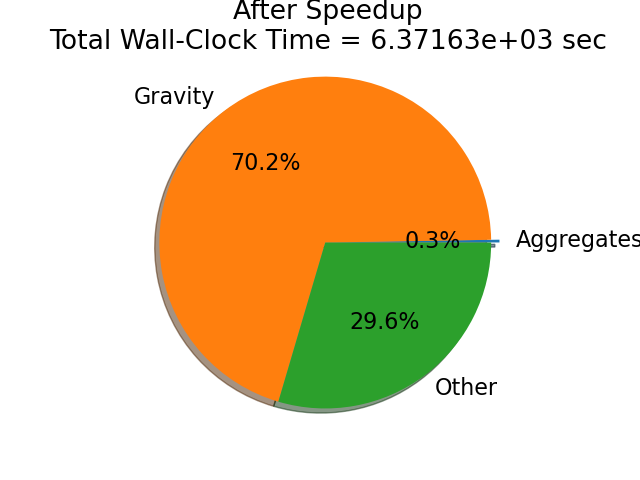}{0.45\textwidth}{(B)}}
\caption{A representative example of profiling results from the interparticle gravity trials described in Section \ref{profiling}, both before (A) and after (B) the efficiency improvements were applied. This test included 20,000 member particles bound into a mix of different aggregate shapes. The set of tests this example was drawn from are described in greater detail in Section \ref{profiling}. The two panels are analogous to the ones shown in Fig.~\ref{fig:profiling-unifgrav}. Note that the second most computationally expensive protocol in this instance before modification is particle self-gravity, in contrast with the example shown in Fig.~\ref{fig:profiling-unifgrav}.}
\label{fig:profiling-selfgrav}
\end{figure}

We compare granular dynamics simulations of particles settling for $10,000$ integration steps run on a single CPU core under both uniform gravity and mutual self gravity and for both the original aggregate routines and the new, more efficient routines for each of the aggregate constructions seen in Fig.~\ref{fig:aggs_demo}. Representative figures (Figs.~\ref{fig:profiling-unifgrav}, \ref{fig:profiling-selfgrav}) compare the profiling metrics for time spent in aggregate routines during the test simulations under the new and old aggregate handling schemes.

\begin{figure}
\gridline{\fig{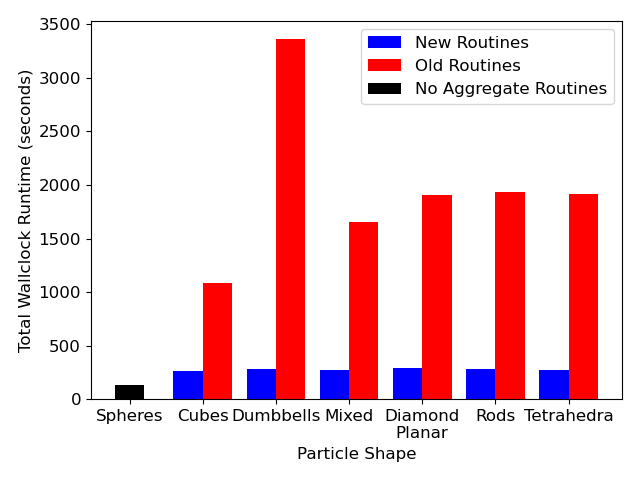}{0.45\textwidth}{(A)}
          \fig{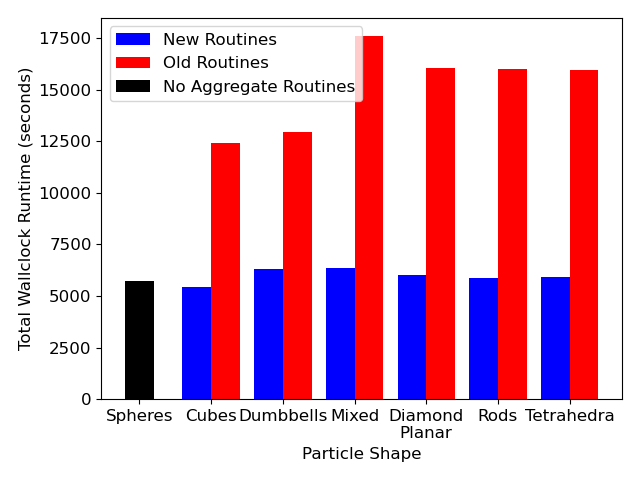}{0.45\textwidth}{(B)}}
\caption{Total wall-clock runtime as a function of aggregate shape for box-filling (A) and collapse (B) simulations run in serial. The blue bars (the left side of the histogram bar pairs) indicate total wall-clock simulation time with the improved search and sorting methods; the red bars (the right side of pairs) indicate total wall-clock simulation time with the original brute-force search. Black bars indicate time spent in analogous free-sphere simulations, with no aggregate options compiled at runtime.}
\label{fig:profiling-tots}
\end{figure}

The profiling simulations of particles in uniform gravity capture a portion of the process of filling a rectangular, open-topped box with bonded aggregates. We use a box with base side lengths of 10 cm and a height of 60 cm. The aggregates are composed of equal-sized, 0.25 cm-radius spheres and we apply lunar-strength gravity (1.62 m s$^{-2}$). For these box-filling simulations, the total number of aggregate member spheres is held constant at $N =$ 6,400 particles, composing between 800 and 3,200 aggregates depending on the aggregate geometry. We also model an analogous scenario with no aggregates but only free spherical particles, as well as a scenario where we mix the symmetric aggregate shapes from Fig.~\ref{fig:aggs_demo} that contain 8 or fewer member particles (but excluding free spheres). In this ``mixed aggregates" scenario, we again keep fixed the particle radii and number of spheres at $N =$ 6,400, consistent with the other box-filling models. Fig.~\ref{fig:profiling-unifgrav} compares the profiling results of the mixed aggregates box-filling tests before and after the efficiency improvements are applied, as a representative for the speedup we see with the updated search and sort routines. The total wall-clock simulation time for $10^4$ time steps in the simulation represented in Fig.~\ref{fig:profiling-unifgrav} is reduced from $1.6\times10^3$ s to $2.7\times10^2$ s---more than five times faster in this case (see Fig.~\ref{fig:profiling-tots} for the direct comparison).

The trial scenarios with interparticle gravity model the gravitational collapse of a cloud of aggregates, which was constructed to ensure that both interparticle contact forces and self-gravity would be at work during the period that we profiled. The collapse simulations use aggregates composed of equal-sized 50 m-radius spheres. For each of the collapse simulations, the number of spherical member particles is held constant at $N = 2 \times 10^4$ particles, ranging from $2.5 \times 10^3$ to $1 \times 10^4$ aggregates, again depending on the number of particles in the aggregate shape. Here, as well, we have analogous models with free spheres and mixed aggregates for comparison. Fig.~\ref{fig:profiling-selfgrav} shows the profiling results for the mixed aggregate collapse as a representative example of the speedup achieved with the updated search and sort methods in a simulation without a uniform gravity field. In this representative simulation, we see a reduction in total wall-clock simulation time over $10^4$ simulation steps from $1.8\times10^4$ s to $6.4\times10^3$ s: almost a factor of 3 decrease in total runtime, as seen in Fig.~\ref{fig:profiling-tots}.

From our profiling, we find that the combined aggregate routines are by far the most expensive operations, easily accounting for over half of the time spent in simulations without our newly updated sorting and searching methods. In analogous tests with the improvements implemented, we find that the total simulation time greatly decreases (see Fig.~\ref{fig:profiling-tots}). Operations on aggregates are now in the noise, representing only a few percent or less of the total simulation time.

To more clearly illustrate the value of the efficiency improvements, we divide the total pre-modification simulation time spent on each test into three segments: the fractions of time respectively devoted to the most costly operation, the second-most costly operation, and all other operations combined (``Other"). We then compare the relative computational costs of these segments before and after the efficiency modifications. We include representative examples for the tests of both uniform gravity (Fig.~\ref{fig:profiling-unifgrav}) and interparticle gravity (Fig.~\ref{fig:profiling-selfgrav}). Prior to the efficiency improvements, aggregate operations accounted for a majority of computation time by a wide margin in both cases. The second-most costly operation was the neighbor search for the uniform gravity tests and gravity calculations in the case of the interparticle gravity tests. In both example trials, the relative cost of the second-most expensive operation compared to the cost of the ``Other" category is constant, as these routines remain unchanged. However, the fraction of runtime spent on aggregate calculations shrinks dramatically, from 84.6\% of total computation time to 2.1\% for the uniform gravity test and from 65.4\% to just 0.3\% for the interparticle gravity test. This corresponds to a substantial decrease in overall runtime. Notably, in simulations using the new implementation, operations on aggregates collectively take more than an order of magnitude less time than the next most expensive operations. The savings in absolute simulation time are substantial as well. We see a typical decrease in total runtime by a factor of 2 to 3 in self-gravity simulations with $N =$ 20,000 particles and by a factor of 4 or more in simulations with a uniform gravity field and $N =$ 6,400 particles (see Fig.~\ref{fig:profiling-tots}). In both sets of models, simulations using the newly updated sort and search methods do not take significantly longer than simulations with the same number of free spheres. Interestingly, we see a dependence on the number of particles in each aggregate that is most pronounced in the box-filling simulations (Fig.~\ref{fig:profiling-tots}A): when using the original routines, simulations with the largest number of aggregates (2-particle dumbbells, with 3,200 aggregates) took significantly longer than simulations with fewer aggregates (8-particle cubes, with 800 aggregates), despite having the same total number of spheres, due to the overhead of repeated brute-force searches. This trend is much less pronounced in the simulations with the new search and sort routines.

We caution that while we can compare models with small aggregates containing the same total number of particles as a simulation with only free spheres, the number of discrete bodies in the aggregate simulations must necessarily be several times smaller than in the spherical simulations and the size of the bodies several times larger. If we want to match the number and size of discrete objects in a simulation with aggregates to a simulation with only free spheres, we would have to replace each sphere with an aggregate of comparable size. Directly replacing spheres with aggregates requires a larger number of total member particles than there were free spheres, because each aggregate must contain several particles, and requires the particles to be smaller than the free spheres so that the total aggregate size matches that of the sphere it is replacing. Including more particles increases the computational load of the gravity calculations (which scale as $O(N\log N)$), and reducing particle size forces smaller time steps in order to resolve particle collisions, as described in \cite{schwartz2012implementation}.

That said, while modeling tens or hundreds of thousands of self-gravitating, non-spherical constituents was previously untenable, these scales are now within reach. To demonstrate this, we conducted additional trials at increasingly high particle resolution. A total of six tests is divided into pairs of spherical particle and mixed aggregate simulations at low, medium, and high resolutions (see Table \ref{tab:large-scale-tests} for labels and results). All six tests consisted of a stationary ``cloud" of constituents collapsing to a settled, stable rubble pile. All initial clouds had nearly the same total mass with the only significant difference between each trial being the constituents. At $\sim$11,000 particles (trials C1 and C2), the wall-clock times approach parity with simulation time. We found that with 2,500 bonded aggregates composed of 10,912 particles (C1), it took 1.20 seconds of wall-clock time to integrate 1 second (or about $2.9 \times 10^{-5}$ dynamical times) of simulation time on 1 CPU core. The spherical counterpart (C2) took 1.50 seconds to run, very similar to the result from C1. For serial runs, performance is typically comparable between aggregate and sphere-based simulations containing the same total number of spheres, as demonstrated in Fig. \ref{fig:profiling-tots}. The largest scale tested used over one million spherical particles, either as free grains (C6) or bound into 250,000 aggregates (C5). One second of integration time required 35.40 seconds of wall-clock time for the sphere-based model and 289.31 seconds for the aggregate model. We also found that tidal disruptions of 10,000 aggregates ($\sim$40,000 particles) could be completed in less than a week on fewer than five CPU cores. These tests were conducted on AMD EPYC 7763 CPU cores using the C compiler included in version 9.4.0 of the GNU Compiler Collection (GCC).

Bonded aggregate performance is comparable to that of spheres in serial, allowing for near real-time simulations for $\sim$2,500 non-spherical constituents. Parallel scaling for aggregates is good enough to allow for high-resolution non-spherical simulations, but still lags behind performance with only spheres. Nominally, we should see $O(N\log N)$ scaling with total particle number $N$. However, limitations inherent in \pkdgrav{}'s architecture prevent us from achieving this ideal. While our optimization efforts have substantially improved these shortcomings, some clear obstacles remain. Currently, we are constrained to store all top-level aggregate information on a single processor when running in parallel, leading to a substantial bottleneck during multi-core simulations. In addition, aggregate operations require all particles to be reordered by aggregate index, which is not required for sphere-based simulations. Since addressing these design limitations would require fundamentally reworking the structure of \pkdgrav{}, we leave this work for future \pkdgrav{} development or for the developers of other codes attempting similar implementations.

\begin{table}
    \centering
    \begin{tabular}{l|l|c|c|c|c}
        \hline
        Trial & Composition & No.\ Constituents & No.\ Spheres & CPU Cores & Wall-Clock Time for 1s Simulation Time\\
        \hline
        \hline
        C1 & Mixed Shapes & 2,500 & 10,912 & 1 & 1.20~s  \\
        C2 & Spheres & 10,912 & 10,912 & 1 & 1.50~s \\
        C3 & Mixed Shapes & 25,000 & 110,150 & 11 & 17.19~s \\
        C4 & Spheres & 110,150 & 110,150 & 11 & 7.51~s \\
        C5 & Mixed Shapes & 250,000 & 1,101,336 & 50 & 289.31~s \\
        C6 & Spheres & 1,101,336 & 1,101,336 & 110 & 35.40~s \\
        \hline
    \end{tabular}
    \caption{A set of six test simulations at increasingly high particle resolutions, with one spherical and one aggregate test at each resolution. Note that to match the constituent resolution of a sphere-based simulation rather than the particle resolution, the corresponding aggregate simulation would need to include \textit{more} spheres than its counterpart, with the ratio depending on the composition of the aggregates. Each aggregate-based trial has the same number of spherical particles as its counterpart. In serial simulations, performance is comparable between sphere-based and aggregate simulations, while parallelized aggregate-based runs become more costly.}
    \label{tab:large-scale-tests}
\end{table}

\section{Applications and future work} \label{sec:applications}

Here we provide three proof-of-concept applications that make use of our approach to soft-sphere contacts with bonded aggregates. The results described here are meant to illustrate the value of the method we describe in this article. These demonstrations are not intended to serve as comprehensive studies, but are part of ongoing work.

\subsection{YORP spinup} \label{subsec:spinup_demo}

Small solar system bodies particularly in the inner solar system are subject to the Yarkovsky– O'Keefe–Radzievskii–Paddack effect (YORP) in which asymmetries in the absorption and re-radiation of solar energy due to surface irregularities result in a small net torque, causing a change in the body's overall spin \citep{rubincam2000radiative}. Over long time spans these small net torques can cause significant spin-ups, potentially reshaping the object \citep{sanchez2012simulation,scheeres2015landslides,cotto2015coupled}, causing it to shed material \citep{sanchez2012simulation,walsh2012spin}, or even form a binary or higher-multiple system \citep{cuk2007formation,walsh2008rotational,jacobson2011dynamics}. 

\begin{figure}
    \centering\includegraphics[width=\linewidth]{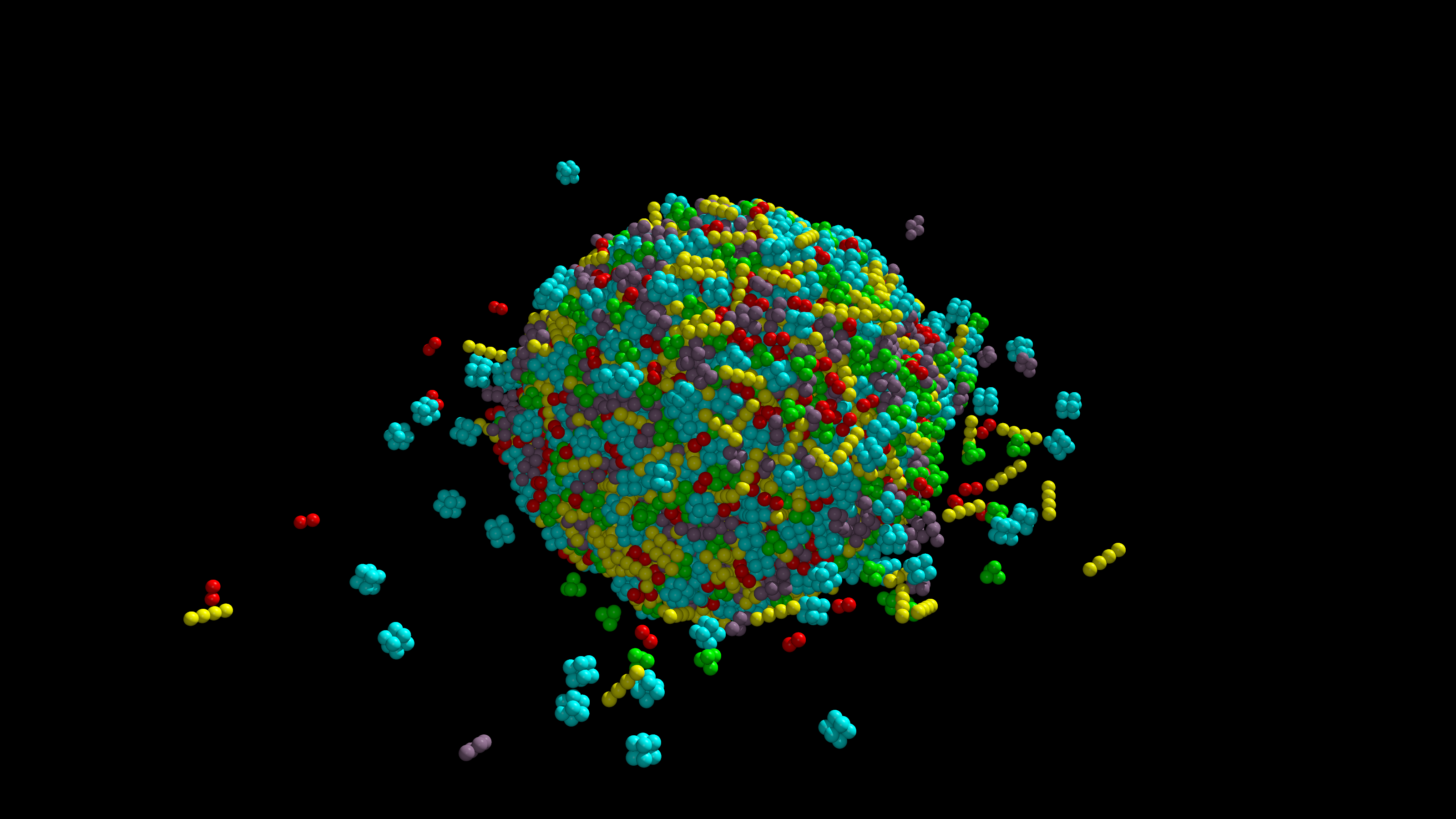}
    \caption{A rubble pile composed of 5,000 bonded aggregates (21,765 particles) shedding mass after being subjected to spinup. Here we use a variety of regular shapes, including cubes, tetrahedra, and 4-particle rods, among others. Aggregates are color-coded by shape.}
    \label{fig:spinup}
\end{figure}

There is much interest in understanding the effects of irregular constituent shape on spin-up outcome, particularly given that some small bodies are such fast rotators that they may require either cohesion or enhanced shear strength (as provided by irregular grain shape, for example) to remain presently stable \citep{rozitis2014cohesive,hirabayashi2014stress,zhang2017creep}. We are modeling spin-up following the approach of \cite{zhang2017creep} by applying an artificial rotational acceleration to a rubble pile, but, unlike in the earlier studies that used independent spheres, we use non-spherical bonded aggregates instead. In contrast with \cite{zhang2017creep}, we apply a constant angular momentum increment at each time step, as opposed to a constant spin period increment. Fig.\ \ref{fig:spinup} shows an example of a rubble pile composed of aggregates losing mass during a spin-up event in \pkdgrav{}.

Our preliminary work indicates that grain shape does have a measurable effect on rubble-pile spinup. For our trial runs, we selected a variety of simple shapes: 2-particle ``dumbbells," 4-particle planar diamonds, 4-particle rod-shaped aggregates, 4-particle tetrahedra, and 8-particle cubes. These shapes are shown in Fig.\ \ref{fig:aggs_demo}. For each shape, we created a progenitor rubble pile composed solely of equal-sized aggregates of that shape. We also included both a rubble pile composed of a mix of all five shapes and four rubble piles composed only of the original, spherical \pkdgrav{} particles. We refer to these runs using the labels S1--S10 (see Table \ref{tab:spinup}). All progenitor bodies were given equal total mass and all constituent spheres had equal mass density. All progenitors contain 20,000 \pkdgrav{} spheres, with the exception of S8 (mixed shapes), which contains 21,764 spheres. Using randomized mixed shapes doesn't allow for an easy division of particles, and we opted instead for the round number of 5,000 aggregates for simpler comparison with other trials. The bulk densities of the settled progenitor rubble piles vary due to inherent differences in packing efficiencies between the different aggregate shapes, so comparing the trials to each other requires some care.

For each settled rubble pile, we adopted a nominal predicted critical spin period $P_{crit} = \sqrt{\frac{3 \pi}{G \rho}}$, computed by equating surface gravity with centrifugal acceleration for a particle at the equator of a rigid, spinning sphere of the given bulk density. We then slowly increased the spin rate of each progenitor until it began to shed material, where we considered a ``disruption" to have occurred when a rubble pile's radius has increased by 5\%. Note that the specific stopping criterion of 5\% used here is somewhat arbitrary; we are interested in the marginal differences in resistance to disruption for each composition, rather than determining a specific disruption threshold for each shape. The observed critical spin period for each composition is recorded as a percentage of the nominal critical spin period calculated for that rubble pile. This allows us to compare trials with different bulk densities to each other directly. We call this metric the ``critical spin ratio." A number greater than 100\% indicates a rubble pile that was easier to disrupt than the nominal case, while a number less than 100\% means the rubble pile was more difficult to disrupt. The results are shown in Table \ref{tab:spinup}. To determine the degree to which our results are dependent on the randomness inherent in the progenitor generation and collapse procedure, we repeated trial S1 once and trial S2 twice with new, randomly generated rubble piles. The two S1 spinups gave critical spin ratios of 102.5\% and 102.6\%, while the three S2 trials gave ratios of 102.6\%, 101.7\%, and 100.3\%. In both cases, the variation is small enough that we feel confident in the conclusions drawn below. Given that we are only presenting a small number of tests for the purpose of a proof of concept, we do not include any further statistical analysis. For both S1 and S2, the average critical spin ratio of all realizations is the number quoted in Table \ref{tab:spinup}.

We have attempted to select test cases that are as similar as possible to allow direct comparison. However, there are some difficulties in engineering a true like-against-like comparison in a test of this nature. Consider trials S1 and S9, for example. Both progenitor rubble piles are composed of 20,000 spherical particles, but in the case of S9 these particles are bound into only 5,000 separate grains, as opposed to 20,000 grains in S1. S1 and S9 have different effective ``resolutions." To account for this, we have included trials S2, S3, and S4 which use progenitor bodies made up of 20,000, 10,000, 5,000, and 2,500 spheres, respectively. We note that in S1--S4 there is an apparent trend of increasing resistance to disruption with increasing resolution, which saturates at around 10,000 particles, though the trials are too limited to draw any firm conclusions. It may be that increased resolution would increase resistance to disruption in the case of other shapes as well. In any case, it is evident from Table \ref{tab:spinup} that using non-spherical shapes increases resistance to spin-up disruption in all cases, regardless of the resolution. The more rounded, ``sphere-like" dumbbells, cubes, and tetrahedra have higher critical spin ratios and so perform more like spheres, though they do confer some degree of additional strength. The case of cubes is more difficult, since they differ by over 10\% in critical spin ratio from the most similar spherical trial (S4), but also score close to 100\%, indicating a critical spin period fairly close to the nominal value. The more elongated diamonds and rods add substantial resistance, with a $\sim$10\% reduction in critical spin ratio compared to S3, and the lowest critical spin ratios of all trials. Trial S8, with mixed shapes, falls in between the extremes.

These early results suggest resistance to breakup in fast-rotating rubble-pile bodies is affected by particle shape, as expected. \cite{walsh2022near} showed that the asteroid Bennu has nearly zero interparticle cohesion, at least near its surface. It may be that particle shape alone could account for the stability of fast rotators without the need to invoke surface cohesion, though a more comprehensive study to establish the typical magnitude of the effect is needed. 

\begin{table}
    \centering
    \begin{tabular}{l|l|c|c}
        \hline
        Trial & Composition & Number of Grains & Critical Spin Ratio \\
        \hline
        \hline
        S1 & Spheres & 20,000 & 102.5 $\pm$ 0.1\% \\
        S2 & Spheres & 10,000 & 101.5 $\pm$ 1.2\% \\
        S3 & Spheres & 5,000 & 104.5\%  \\
        S4 & Spheres & 2,500 & 108.6\%  \\
        S5 & Dumbbells & 10,000 & 98.4\% \\
        S6 & Cubes & 2,500 & 98.2\% \\
        S7 & Tetrahedra & 5,000 & 97.8\% \\
        S8 & Mixed Shapes & 5,000 & 96.8\% \\
        S9 & Diamonds & 5,000 & 96.3\% \\
        S10 & Rods & 5,000 & 94.0\% \\
        \hline
    \end{tabular}
    \caption{Observed critical spin as a fraction of the nominal predicted value (see Section \ref{subsec:spinup_demo}). A rubble pile composed of many, only-spherical particles (S1) nearly matches the predicted value, while a body made up of rod-shaped grains deviates most strongly. All progenitor bodies have approximately equal total mass and bulk radius.}
    \label{tab:spinup}
\end{table}

\subsection{Tidal encounters}\label{tidal_demo}

\begin{figure} 
\gridline{\fig{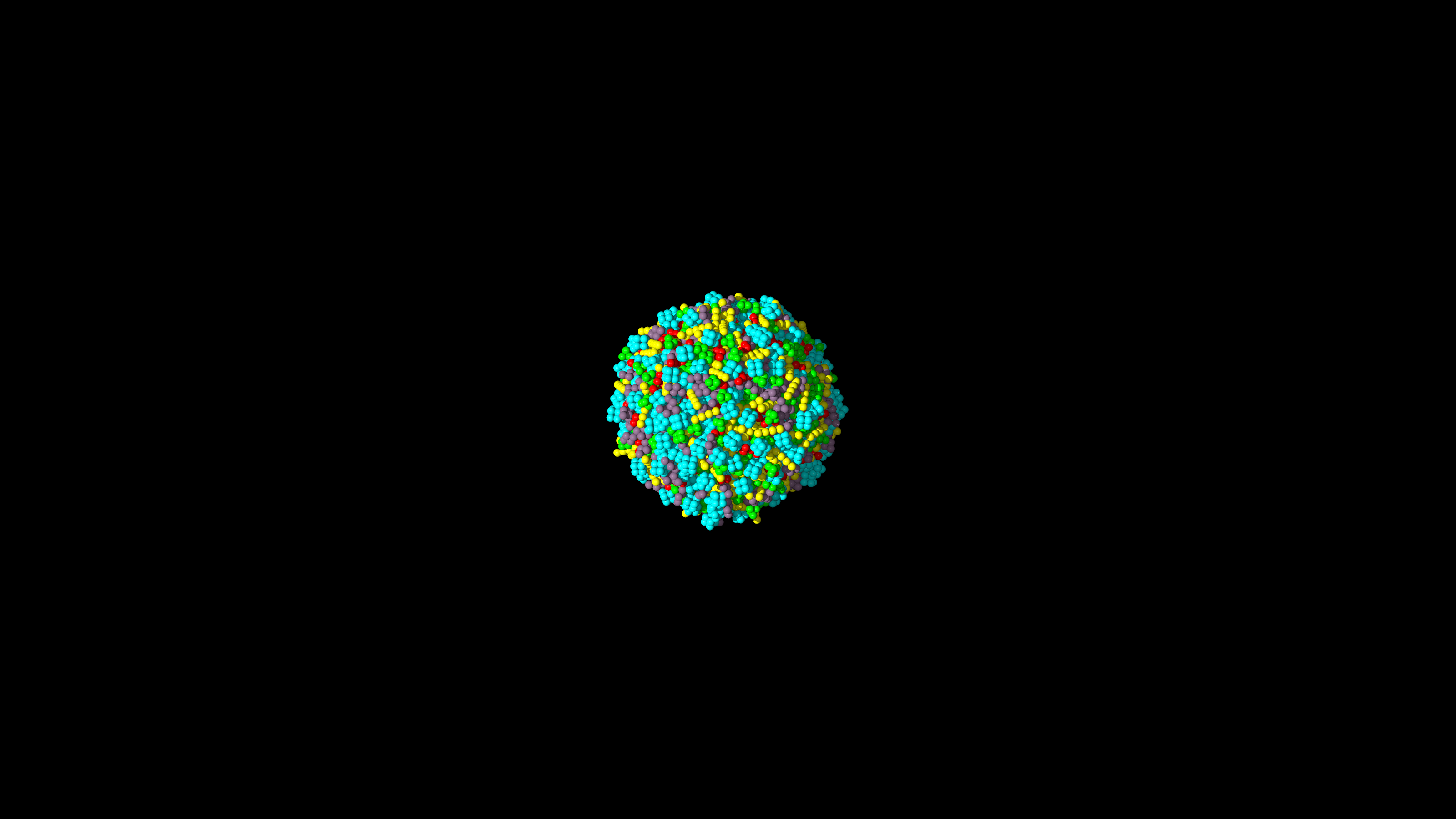}{0.3\textwidth}{(A)}
          \fig{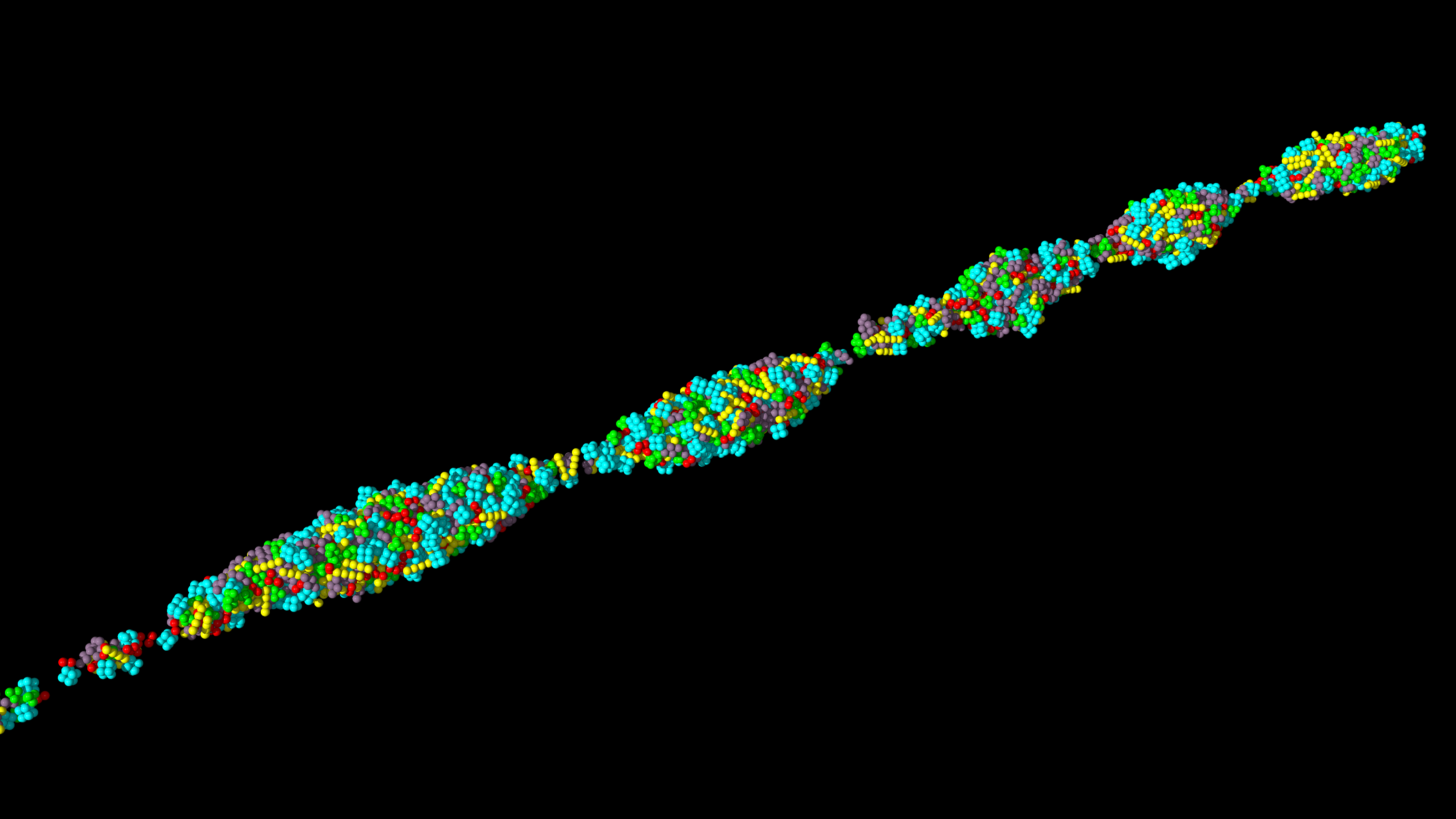}{0.3\textwidth}{(B)}
          \fig{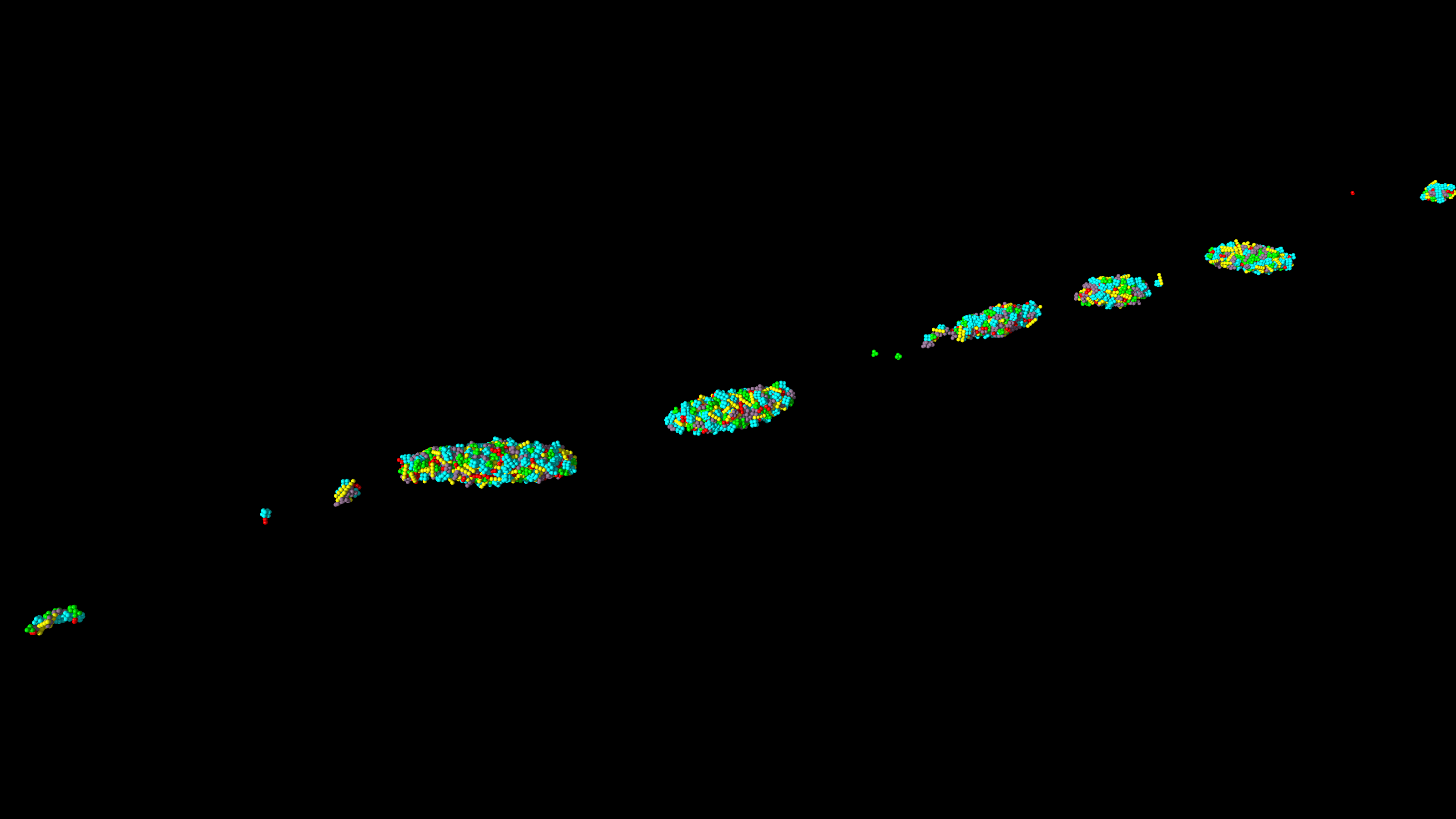}{0.3\textwidth}{(C)}}
\caption{A rubble pile body composed of 5,000 bonded aggregates (21,765 particles) before and after a tidal encounter. As in Fig.\ \ref{fig:spinup}, aggregates are color-coded by shape.}
\label{fig:tidal}
\end{figure}

When small solar system bodies like asteroids and comets pass near large, dense bodies, they can be subject to significant tidal forces. Depending on the specific parameters of the encounter, these tidal forces can cause small disturbances \citep{richardson1998tidal,yu2014numerical,demartini2019using}, resurfacing \citep{binzel2010earth,yu2014numerical}, changes in spin state \citep{scheeres2004evolution,scheeres2005abrupt}, and even reshaping or disruption of the entire body \citep{bottke19991620,walsh2006binary,zhang2020tidal}. Some have speculated that the interstellar object `Oumuamua's unusual elongated shape could be the result of the same tidal encounter that ejected it from its host system \citep{cuk20181i,zhang2020oumuamua}. It is plausible that the additional internal friction conferred by non-spherical particles could play a significant role in determining its very elongated shape after a tidal encounter, although \cite{zhang2020oumuamua} instead invoke sintering of bonds between surface particles as a possible explanation. While recent work by \cite{zhang2020tidal} applied spherical-particle version of \pkdgrav{} to the question of tidal distortion of rubble piles, the role of particle shape in this process has not been investigated. See Fig.\ \ref{fig:tidal} for an illustration of a tidal disruption simulation using non-spherical particles in \pkdgrav{}.

We conducted a set of ten trials (T1--T10) to assess the effects of particle shape on tidal reshaping and disruption. The progenitor rubble piles are identical to those used in Section \ref{subsec:spinup_demo} and are named accordingly, so the rubble pile in trial T1 is the same as the rubble pile in trial S1, and so on. In each trial, a rubble pile passes by an Earth-mass particle with a close-approach distance of 1.2 Earth radii and is disrupted by tidal forces. For each trial, we set a nominal disruption threshold distance using the Roche limit $d \approx 1.26 R \sqrt[3]{\rho_M/\rho_m}$, where $R$ is the radius of the primary (disrupting) body, $\rho_M$ is the density of the primary, and $\rho_m$ is the bulk density of the rubble pile. This value is not intended to serve as a ``prediction" of the disruption threshold. In analogy with Section \ref{subsec:spinup_demo}, we use it as a normalization factor to allow direct comparison of disruption thresholds for different trials with different bulk densities. For each trial, we record the distance of the rubble pile from the primary when the rubble pile's bulk radius has increased by 0.5\%\footnote{As in Section\ \ref{subsec:spinup_demo}, the 0.5\% disruption threshold figure is arbitrary. In this case, it was chosen in order to produce a disruption ratio near 100\% for trial T1, recognizing that these are not fluid bodies and therefore we would expect them to disrupt interior to the normalization distance. Threshold values of 1\% and 5\% produced comparable results.} of its initial value. We then quote this distance relative to the nominal disruption threshold calculated for that progenitor body as a percentage. We refer to this value as the ``disruption ratio." Higher values indicate lower resistance to tidal disruption. Results for each trial are shown in Table \ref{tab:tidal}. We again conducted two realizations of T1 and three realizations of T2 to determine how resistant our results were to random variation. The disruption ratios were 83.8\% and 86.8\% for the T1 trials and 77.0\%, 78.4\%, and 79.1\% for the T2 trials. As in the case of Section \ref{subsec:spinup_demo}, we feel that the natural variation here is small enough not to affect our conclusions. For T1 and T2, we quote the average disruption ratio across all realizations.

In contrast to Section \ref{subsec:spinup_demo}, we see a trend toward decreasing resistance to disruption with increasing resolution when comparing runs T1--T4. This is reasonable since, near the disruption threshold, the smaller constituents in the higher-resolution simulations will be able slide past each other under shear forces more easily than larger constituents will. The difference in results between spheres (T1--T4) and bonded aggregates (T5--T10) is more varied in the case of tidal disruption as it was in the case of spin-up disruption. When comparing runs of the same resolution the spherical progenitors are always easier to disrupt, with the exception of dumbbells. Rods (T10) show the greatest change in disruption threshold when compared to their spherical analogs, followed by planar diamonds (T9). By this metric, rod-shaped and diamonds-shaped aggregates are again the most difficult to disrupt. This is reasonable, given their highly non-spherical shapes. The overall tendency again is toward greater resistance for progenitors composed of non-spherical shapes. As in the case of spinup, a more in-depth study is needed to fully characterize and quantify the effect.

\begin{table}
\centering    
    \begin{tabular}{l|l|c|c}
        \hline
        Trial & Composition & Number of Grains & Disruption Ratio \\
        \hline
        \hline
        T1 & Spheres & 20,000 & 85.3 $\pm$ 1.5\% \\
        T2 & Spheres & 10,000 & 78.2 $\pm$ 1.2\% \\
        T3 & Spheres & 5,000 & 74.7\% \\
        T4 & Spheres & 2,500 & 63.8\% \\
        T5 & Dumbbells & 10,000 & 78.7\% \\
        T6 & Cubes & 2,500 & 54.9\% \\
        T7 & Tetrahedra & 5,000 & 62.7\% \\
        T8 & Mixed Shapes & 5,000 & 70.1\% \\
        T9 & Diamonds & 5,000 & 60.8\% \\
        T10 & Rods & 5,000 & 58.6\% \\
        \hline
    \end{tabular}
    \caption{Observed disruption threshold distance as a fraction of the Roche limit for each tidal disruption trial (see Section \ref{tidal_demo}). With the exception of T5, trials with bonded aggregates show more resistance to disruption than the analogous trials using spherical particles.}
    \label{tab:tidal}
\end{table}

\subsection{Brazil-Nut effect}

The Brazil-nut Effect (BNE) \citep{rosato1987brazil} is a suggested mechanism for the vertical migration of boulders in a granular medium, in which frictional interactions between particles permit larger blocks to rise to the surface when subjected to repeated seismic shaking. Granular convection, where constituents move down along confining walls and up through the central medium, is the dominant mechanism in confined experiments \citep{asphaug2007asteroid}, but simulations using periodic boundary conditions find that small grains moving into empty areas below the large constituents (“void-filling”) plays a larger role in low-gravity BNE models \citep{maurel2016numerical, chujo2018categorization}. Both convection and void-filling are influenced by friction: when interactions disrupt the flow of grains past one another, their ability to convect or move into voids is disrupted, thus halting the rise of the ``Brazil nut'' (large intruder). Furthermore, we expect to find more voids in systems of irregularly shaped grains due to the increased shear strength of aggregates versus their spherical counterparts, leading to a more porous equilibrium state when packing under uniform gravity. The BNE has been investigated several times in the past with \pkdgrav{} \citep{matsumura2014brazil, maurel2016numerical, chujo2018categorization} and recent studies have shown how ellipsoidal shapes influence the process of boulder stranding \citep{zeng2022new}. However, no investigation has yet been published using the varied geometries that we can model with our approach. The non-spherical BNE could also explain the highly porous, boulder-dominated surfaces of rubble piles like Bennu, the target of the recent OSIRIS-REx sample return mission \citep{walsh2022near}. 

\begin{table}
    \centering    
    \begin{tabular}{l|l|l|c|c}
        \hline
        Trial & Composition & Intruder Shape & Number of Grains & Rise Time (cycles)\\
        \hline
        \hline
        B1 & Dumbbells & Sphere & 2,000 & 18 \\
        B2 & Tetrahedra & Sphere & 1,000 & 13 \\
        B3 & Diamonds & Sphere & 1,000 & 28 \\
        B4 & Rods & Sphere & 1,000 & 22 \\
        B5 & Cubes & Sphere & 500 & -- \\
        B6 & Spheres & Sphere & 4,000 & 36 \\
        B7 & Mixed Shapes & Axisymmetric Aggregate & 900 & 32 \\
        \hline
    \end{tabular}
    \caption{Observed rise time of a large intruder grain in a medium of aggregates (B1--B5, B7) or spheres (B6). The intruder notably rises faster in media made of irregularly shaped grains in all cases except for B5, where it does not rise at all, likely due to high packing efficiency in the medium. All particles have equal density but the intruder is the largest constituent in radius.}
    \label{tab:BNE}
\end{table}

\begin{figure}
\centering
\gridline{\fig{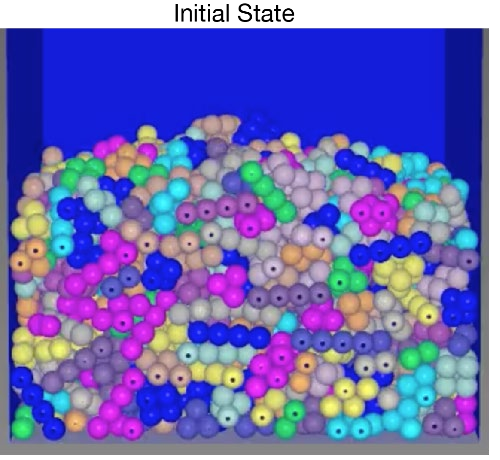}{0.4\textwidth}{(A)}
          \fig{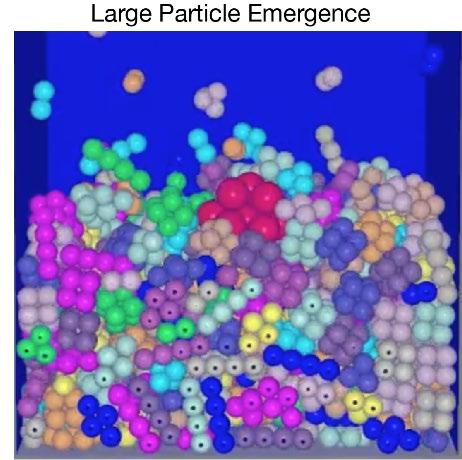}{0.4\textwidth}{(B)}}
\gridline{\fig{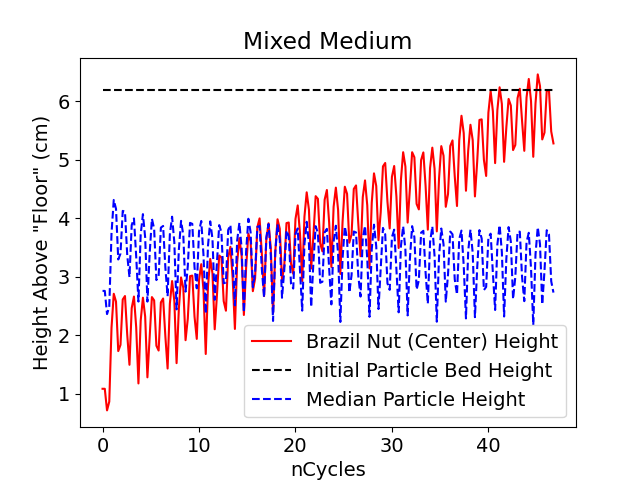}{0.5\textwidth}{(C)}}
          \caption{Images of the initial packing of a system of mixed aggregates (color-coded by shape) with a large intruder buried in the medium (A) and the intruder (red) emerging after repeated sinusoidal shaking of the medium (B). Panel (C) is a plot demonstrating rise of the intruder particle and thus the non-spherical Brazil-nut Effect. In panel (C), the red, solid line demonstrates the height of the center of the intruder particle, the blue, dashed line indicates the median height of the non-intruder particles in the medium, and the black, dashed line indicates the initial height of the particle bed.}
          \label{fig:bne}
\end{figure}

\begin{figure}
    \centering
    \includegraphics[width=\textwidth]{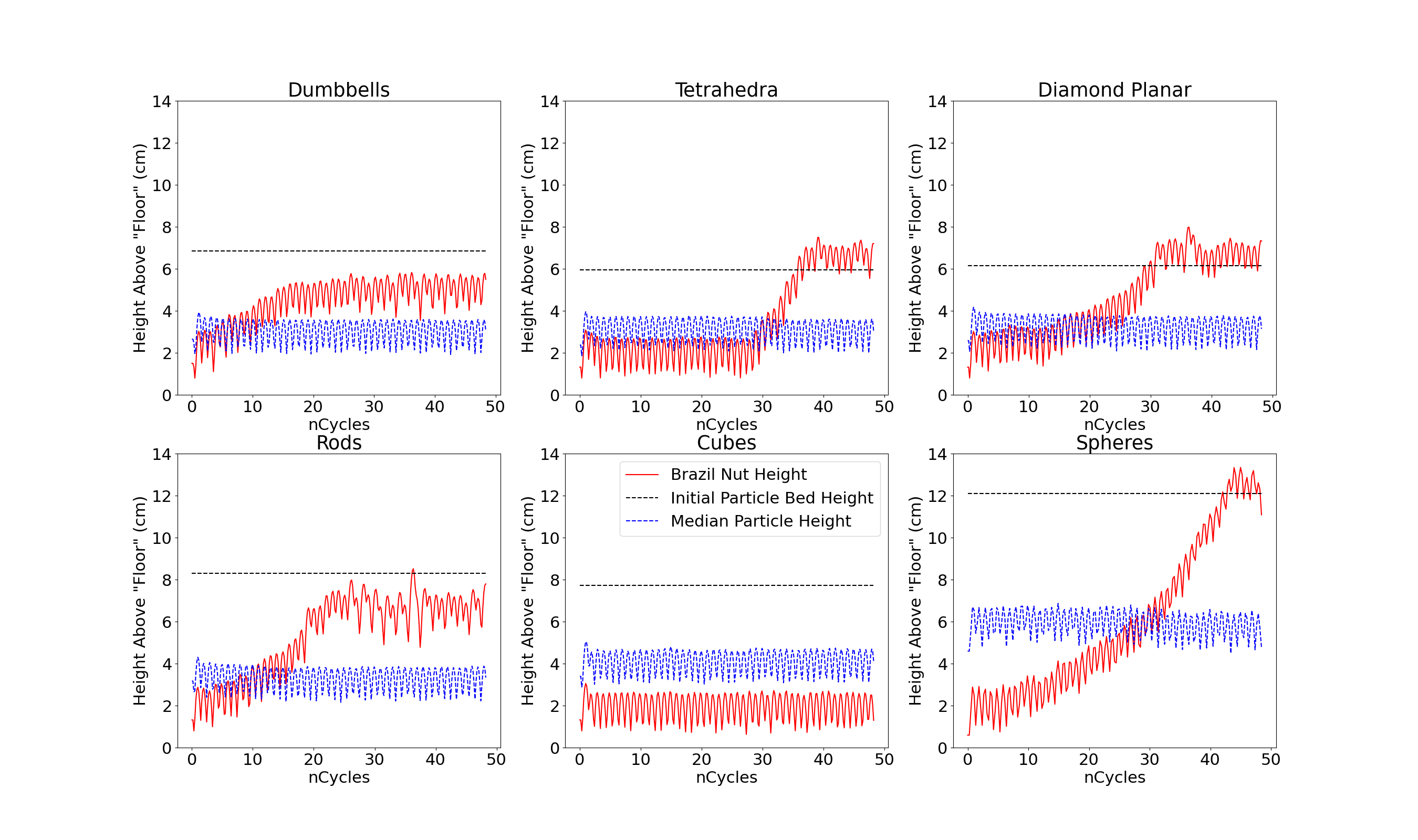}
    \caption{A grid of plots demonstrating the rise of the intruder particle surrounded by media of different symmetric aggregate shapes, each with the same curves and definitions as described in Fig.~\ref{fig:bne}C. From left to right and top to bottom, the media around the intruders are composed of: 2-particle dumbbell shapes, 4-particle tetrahedra, 4-particle planar diamond shapes, 4-particle rods, 8-particle cubes, and free spheres.}
    \label{fig:bne_panel}
\end{figure}

We conducted a set of 7 sample trials (B1--B7) to assess the effects of particle shape on the rise speed of large intruder constituents in a medium of irregularly shaped grains. We initialize our system for these trials by placing a large spherical (for B1--B6) or aggregate (B7) intruder at the bottom of a rectangular chamber of base area $10 \times 10$ cm$^2$ under uniform Earth gravity ($a_g = -9.8$ m s$^{-2}$). We then fill the chamber with grains of the shape defined in Table~\ref{tab:BNE} such that the total number of particles is about 4,000. The intruder has radius 1.5 cm and the radius of aggregate component particles (or free spheres in B6) is 0.25 cm but all particles in the simulation have an equivalent density of 2.7 g cm$^{-3}$ and the same gravel-like friction parameters \citep{zhang2017creep}. Once we have filled the chamber, we force a sinusoidal oscillation of the walls in the system, with oscillation amplitude 1 cm and frequency 54.2 rad sec$^{-1}$, giving the system a dimensionless acceleration (Froude Number; see \cite{matsumura2014brazil}) of about 3. We shake the system for 50 cycles and track the rise of the large intruder through the medium, as seen in Figs.~\ref{fig:bne} and \ref{fig:bne_panel}, with results shown in Table~\ref{tab:BNE}.

Fig.~\ref{fig:bne} shows the initial frame and the frame of intruder emergence from trial B7, a sample model of the BNE with a mix of irregular aggregate shapes in the medium around a larger intruder aggregate. The third frame in Fig.~\ref{fig:bne} shows the height of the center of mass of the intruder (red), the median center of mass height for non-intruder aggregates (blue) and the initial bed height (black). Similar rise plots are shown for trials B1--B6 in Fig.~\ref{fig:bne_panel}. What we can see from these plots and Table~\ref{tab:BNE} is that in almost every case tested, the speed of the intruder's rise, measured by inspection from when it moves above its average initial height (averaged over the oscillations) at the bottom of the container to when it reaches its final height (again, averaged over the oscillations) at the top, is faster with a medium made of irregularly-shaped aggregates than in one composed of only spheres. An interesting exception comes in the case of a medium constructed of cube shaped aggregates (B5), which suppresses the rise of the Brazil nut. This may be due to the relative sizes of the cubes to the intruder (an average ratio of $1:\sqrt{3}$) reducing the efficiency of void filling. It could also be that a medium constructed of only cubes has a significantly higher effective shear strength than the other aggregate media we tested, and thus requires more vigorous shaking to move the intruder up through the medium. All of these models require further work, including testing with periodic lateral boundaries \citep{maurel2016numerical} and investigations with different and potentially more realistic shaking waveforms. However, the initial results shown here indicate that modeling with irregularly-shaped grains can lead to a systematically faster rise of large subsurface blocks \citep{demartini2020efficient}.

\section{Conclusion} \label{sec:conclusion}

Grain shape is known to play an important role in determining shear strength and resistance to deformation in the context of granular media \citep{wegner2014effects}. However, the effects of non-spherical grains have not been considered in most numerical work on rubble-pile bodies to date or have at best been restricted to limited, low-resolution studies. We have presented an implementation of non-spherical grains of arbitrary shape with soft-sphere contacts in our DEM $N$-body code \pkdgrav{}. As far as we are aware, \pkdgrav{} is the only code that combines an $N$-body gravity solver and a soft-sphere contact model with an efficient “glued-sphere” approach to constructing non-spherical constituents. \pkdgrav{}'s existing optimizations and parallel implementation allow us to conduct simulations of self-gravitating, colliding, irregular grains complete with gravitational and contact torques, even when including hundreds of thousands of constituents. Preliminary tests of our method, applied to YORP spinup, tidal disruption, and the Brazil-nut effect, show the potential importance of simulations that include non-spherical grains. Grain shape may be a factor in determining the resistance to disruption of rubble-pile bodies subject to planetary tides or spinup and could help explain efficient boulder stranding on bodies with regolith surfaces via the Brazil-nut effect, and these effects are evident in our trial simulations. Our method improves realism while retaining high dynamic range and efficiency.

\section{Acknowledgments}

This work was partially supported by FINESST grants 80NSSC20K1392 and 80NSSC21K1531 awarded by the National Aeronautics and Space Administration (NASA). This material is based in part upon work supported by the National Science Foundation under Grant No.\ 2108441 and work supported by NASA under Grant No.\ 80NSSC19M0216 issued through the Solar System Exploration Research Virtual Institute. Simulations were performed on the University of Maryland's Deepthought2 and Zaratan clusters, as well as the YORP cluster hosted by the University of Maryland Department of Astronomy. Some visualizations use the Persistence of Vision Ray Tracer (\texttt{www.povray.org}), and analysis was carried out in part using the NumPy and SciPy python modules.

\bibliography{sources.bib}
\bibliographystyle{aasjournal}

\appendix \label{app:AggFuncTable}
\begin{table}[ht]
\centering
\caption{Functions modified by updates in Section \ref{search}, each of which previously contained a brute-force search for aggregate members.}
\begin{tabular}{ l | l }
 \hline
 Function Basename & Definition \\
 \hline \hline
 CountPart & Counts number of particles in an aggregate\\
 GetAccelAndTorque & Returns center-of-mass (COM) acceleration of and torques\\
  & \hspace{5mm} on an aggregate\\
 GetAxesAndSpin & Builds inertia tensor, returns angular momentum vector\\
  & \hspace{5mm} relative to COM\\
 GetCOM & Returns COM position, velocity, and total mass of aggregate\\
 SetBodyPos & Transforms positions of aggregate particles to body frame\\
 SetMassDEM & Stores aggregate total mass in constituent particle structures,\\
  & \hspace{5mm} needed for SSDEM routines\\
 SetSpacePos & Transforms positions of aggregate particles to space frame\\
 SetSpaceSpin & Transforms spins of aggregate particles to space frame\\
 SetSpaceVel & Transforms velocities of aggregate particles to space frame\\
 \hline
\end{tabular}
\label{tab:funcs}
\end{table}

\end{document}